\documentclass[a4paper,11pt]{article}
\usepackage{jinstpub} 
\usepackage{lineno}
\usepackage{caption}
\usepackage{subcaption}

\title{\boldmath Contribution of the light-collection non-uniformity to the energy resolution for the spaghetti-type calorimeter modules}

\author[a,b]{V. Guliaeva,}

\author[c]{S. Kholodenko,}

\author[a]{E. Shmanin}

\author[a,b]{and A. Anokhina}

\affiliation[a]{National University of Science and Technology <<MISIS>>,\\ 119049, Leninskiy Prospekt 4, Moscow, Russia}

\affiliation[b]{Moscow State University <<MSU>>\\ 119991, GSP-1, 1-2 Leninskiye Gory, Moscow, Russia}

\affiliation[c]{Istituto Nazionale di Fisica Nucleare Sezione di Pisa,\\
3 Largo Bruno Pontecorvo, Pisa, Italy}

\emailAdd{vasilisa.guliaeva@cern.ch}

\abstract{

Spaghetti-type calorimeters (SpaCal) are being considered as a potential solution for experiments at the High-Luminosity Large Hadron Collider (HL-LHC).

SpaCal modules consist of an absorber block with a matrix of holes filled with scintillating fibres. This geometry offers a flexible granularity. However, the total number of scintillating fibres per channel could exceed the available photocathode surface area, necessitating a light guide to efficiently collect and register the scintillating light from the scintillating fibres to a photomultiplier. The non-uniformities in the light collection would impact the energy resolution of the detector.

In this study, the impact of the light collection non-uniformity on the energy resolution is estimated for the various geometries of light guides using optical simulations~(GEANT4 simulation with optical photons). The study performed assumed a SpaCal prototype with $30\times 30$~mm$^2$ cell size and photomultipliers featuring realistic entrance window shapes and sizes, and explores different light guide designs.}


\begin{document}
\maketitle
\flushbottom

\section{Introduction}
\label{sec:intro}

Spaghetti-type~(SpaCal) calorimeters are well-known since the early 1990s~\cite{Scheel} and have impressive performance in terms of energy resolution~\cite{Armstrong:1998qs}.
The main advantages are the flexibility to define and change the detector granularity, as well as the reduction of the shower’s transverse size by using dense materials as an absorber. Nevertheless, there is an open question of the uniformity of the light collection and how it affects the energy resolution.

This work investigates the impact of light collection non-uniformity on the energy resolution of SpaCal modules. As a prototype, we consider the design of a cast lead SpaCal module developed in NUST MISIS for the LHCb ECAL Upgrade Phase 2~\cite{Kholodenko:2022anr}. The details on the single cell design and material optimisations are presented in ~\cite{Shmanin:2023oqo}. We assume that the test prototype is inclined by 3 degrees relative to the beam direction, as described in the LHCb Particle Identification Enhancement Technical Design Report~\cite{Lindner:2866493}.

To evaluate the energy resolution, the amount of energy deposited in each scintillating fibre is weighted by an additional coefficient, which emulates the light collection non-uniformity. These additional coefficients are obtained either using a Gaussian distribution with a variable sigma~(Section~\ref{sec:GaussianApprox}) or directly from detailed optical simulations of specific light guide designs~(Section~\ref{sec:EfficiencyLG}). For the last option, optical simulations were developed to estimate the effect of different light guide geometries.

\section{Simulation description}
\label{sec:Calo}

The simulation was performed using the GEANT4 toolkit to analyze the impact of different light guide geometries on light collection uniformity ~\cite{GEANT4:2002zbu,Allison:2006ve,Allison:2016lfl}.  The GEANT4 11.1 version is used with the FTFP\_BERT physics list, including the G4OpticalPhysics class, to simulate the generation and propagation of optical photons in materials. 

\begin{figure}[htbp]
\centering
\includegraphics[width=.8\textwidth]{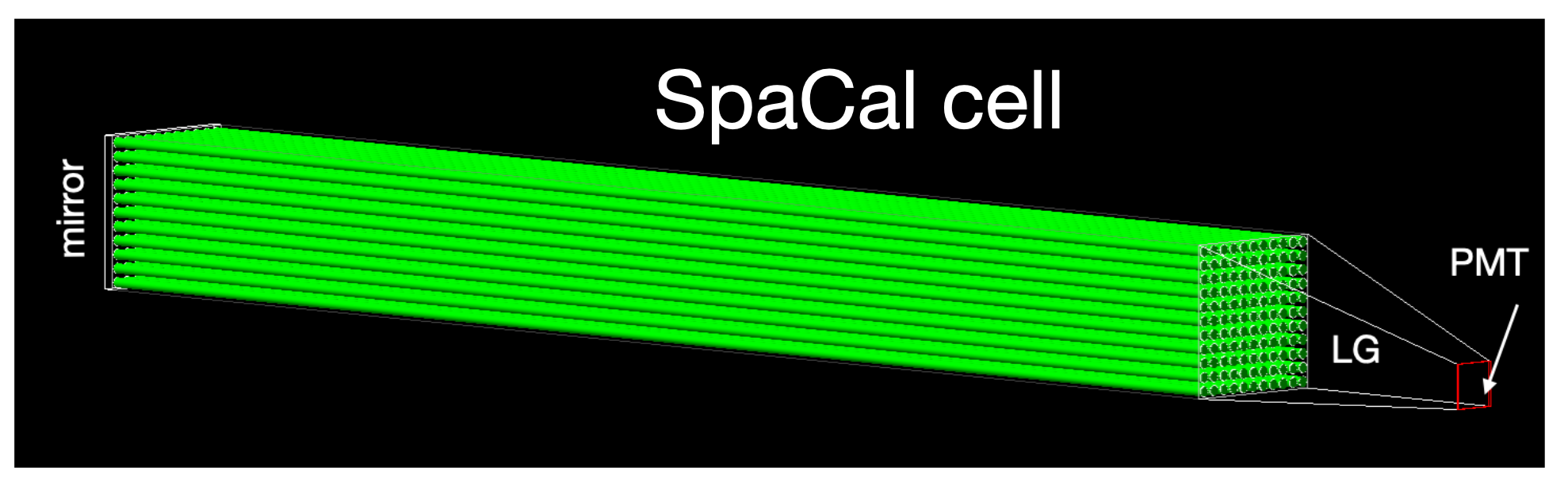}
\caption{Geometry rendering of the SpaCal Cell Prototype with light guide (LG) and photomultiplier tube (PMT) Setup. \label{fig:SpaCal_cell}}
\end{figure}

The single-cell SpaCal prototype has a transverse size of $30.25\times30.25$~mm$^2$ and a length of 300~mm (Figure~\ref{fig:SpaCal_cell}). The model comprises a lead absorber block with an $11\times 11$ matrix of holes. Each hole is equipped with a steel capillary tube and scintillating fibre. The steel tubes provide structural support and housing for the fibres. Each tube has an inner diameter of 2.2~mm and an outer diameter of 2.4~mm with a spacing of 2.75~mm between the centers of adjacent tubes. The scintillating fibres are $\oslash $ 2~mm, featuring the properties such as light yield, attenuation length and emission spectra specified for the SCSF-78 scintillating fibres, with single cladding designed using appropriate refractive indices~\cite{kuraray}.

\begin{figure}[htbp]
\centering
\includegraphics[width=.5\textwidth]{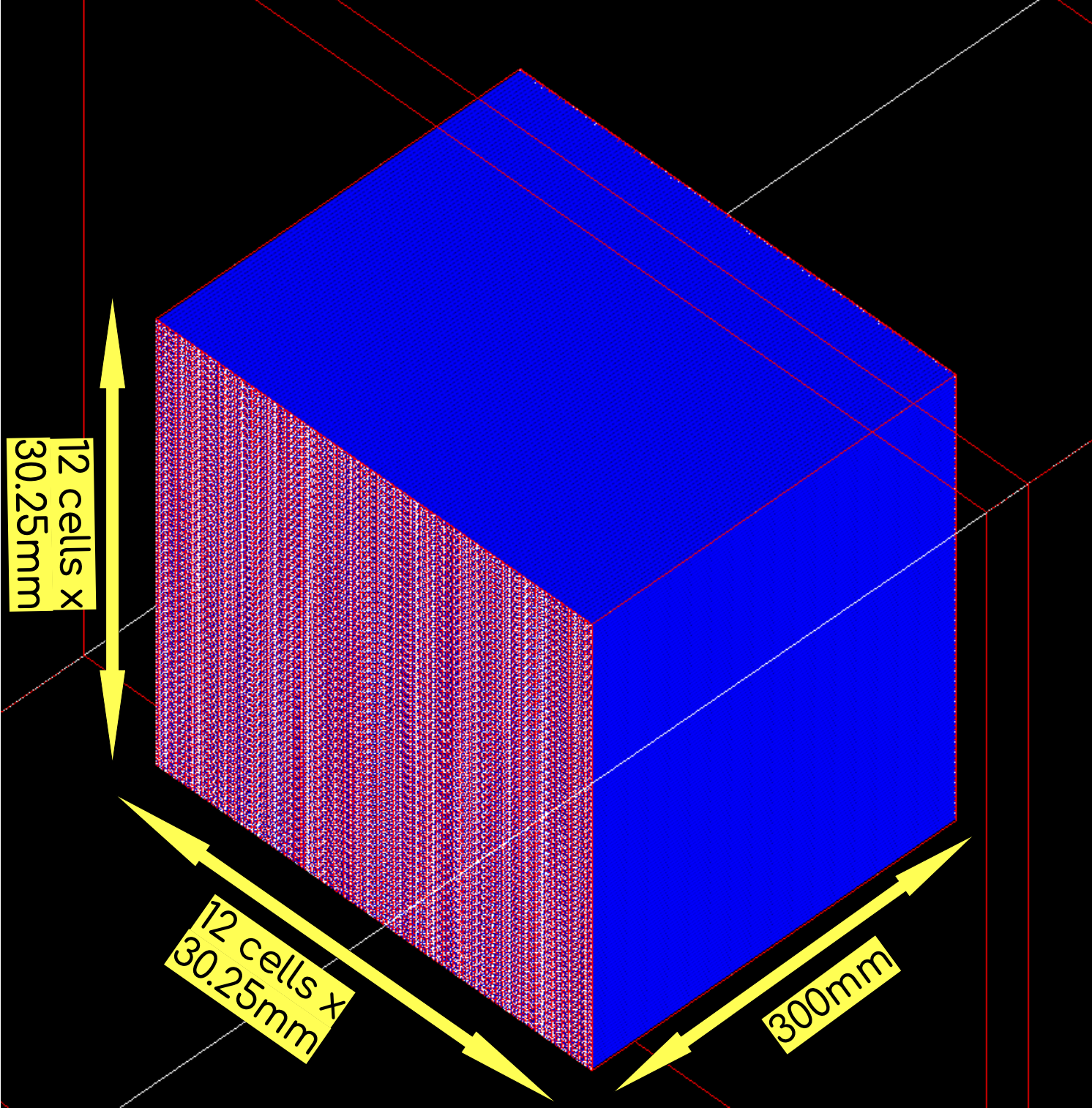}
\caption{GEANT4 simulation of the SpaCal module. \label{fig:SPACAL_wholemodule}}
\end{figure}

The test prototype (the SpaCal module) for the simulation was constructed by arranging a set of $12 \times 12$ single cells in both the X and Y directions, forming a transverse section of $363 \times 363$~mm$^2$, while maintaining the same length of 300~mm for each cell (see Figure~\ref{fig:SPACAL_wholemodule}). Each cell is equipped with a dedicated light guide.

\begin{figure}[htbp]
\centering
    \begin{minipage}{0.48\linewidth}
    \includegraphics[width=\linewidth]{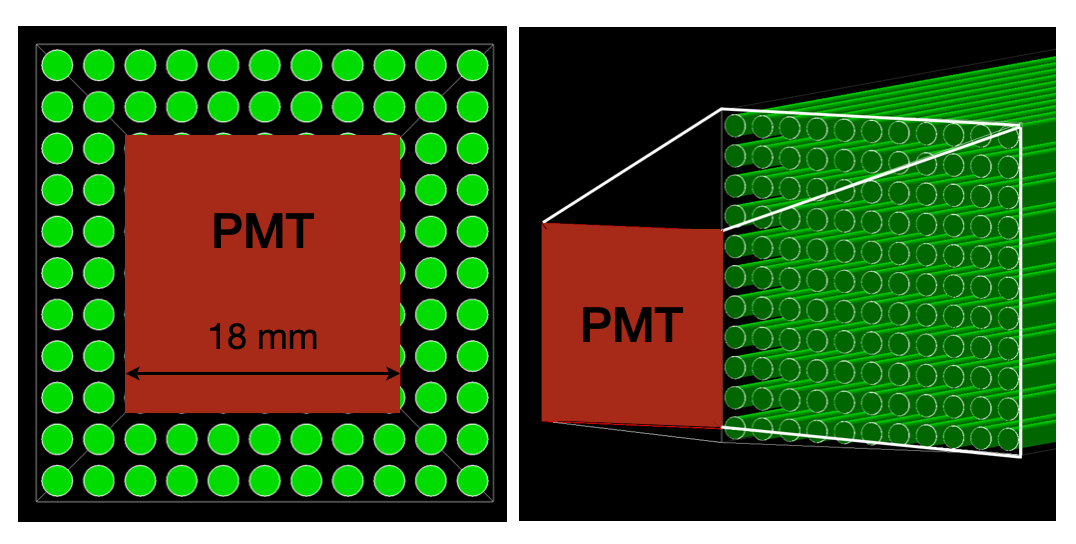}
    \caption*{Geometry 1}
    \end{minipage}
    \begin{minipage}{0.48\linewidth}
    \includegraphics[width=\linewidth]{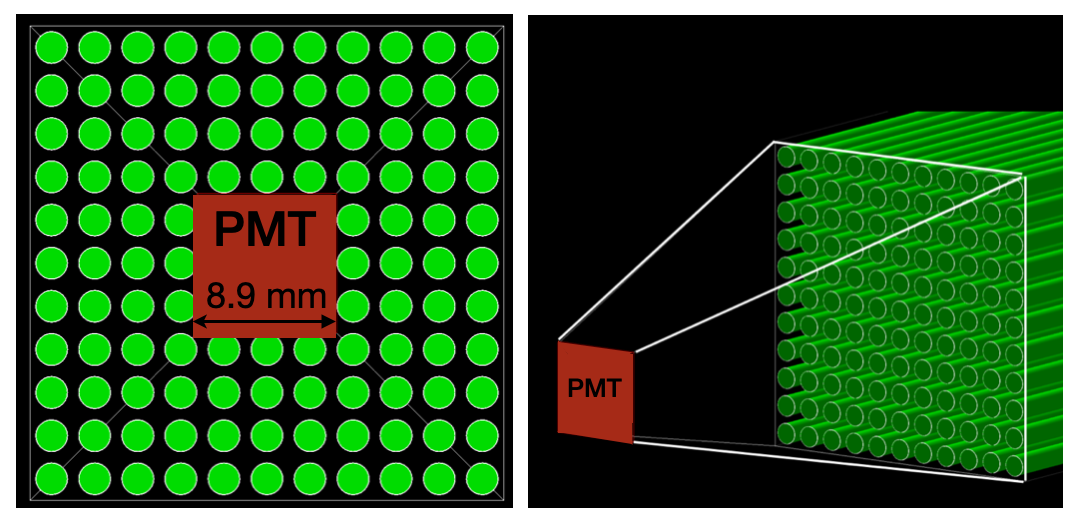}
    \caption*{Geometry 2}    
    \end{minipage}
    \vspace{10mm}
    \begin{minipage}{0.48\linewidth}
    \includegraphics[width=\linewidth]{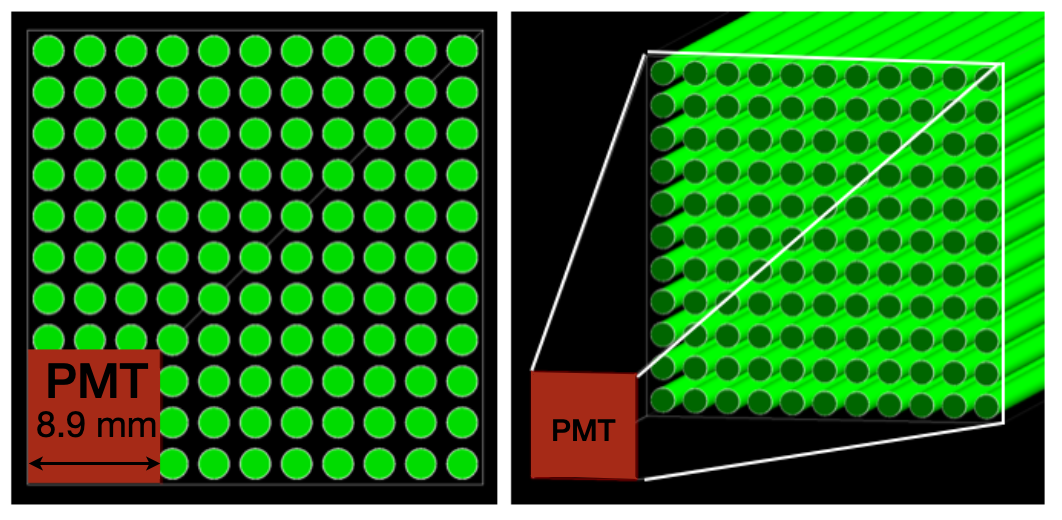}
    \caption*{Geometry 3}    
    \end{minipage}
    \begin{minipage}{0.48\linewidth}
    \includegraphics[width=\linewidth]{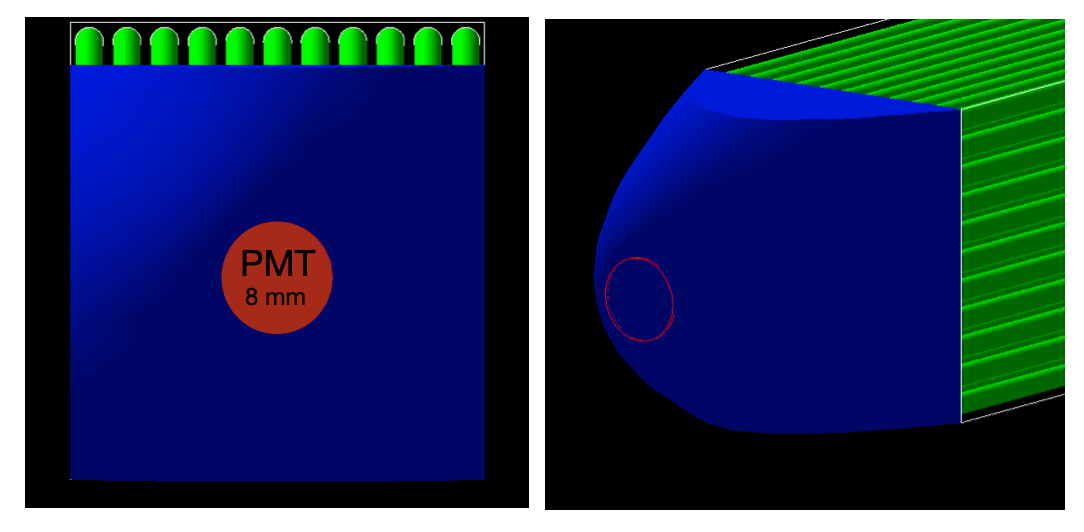}
    \caption*{Geometry WC}    
    \end{minipage}

\caption{Schematic view presenting the studied LG geometries. Geometry 1: Symmetrical trapezoid with $18 \times 18$~mm$^2$ PMT base;
Geometry 2: Symmetrical trapezoid with $8.9 \times 8.9$~mm$^2$ PMT base;
Geometry 3: Asymmetrical trapezoid with $8.9 \times 8.9$~mm$^2$ PMT base located in the corner. This option represents the use of the multi-anode PMT to read four cells simultaneously.
Geometry WC: Winston cone with round $\oslash 8$~mm PMT base.} 
\label{fig:LGGeo}
\end{figure}

\begin{figure}[htbp]
\centering
\includegraphics[width=.28\textwidth]{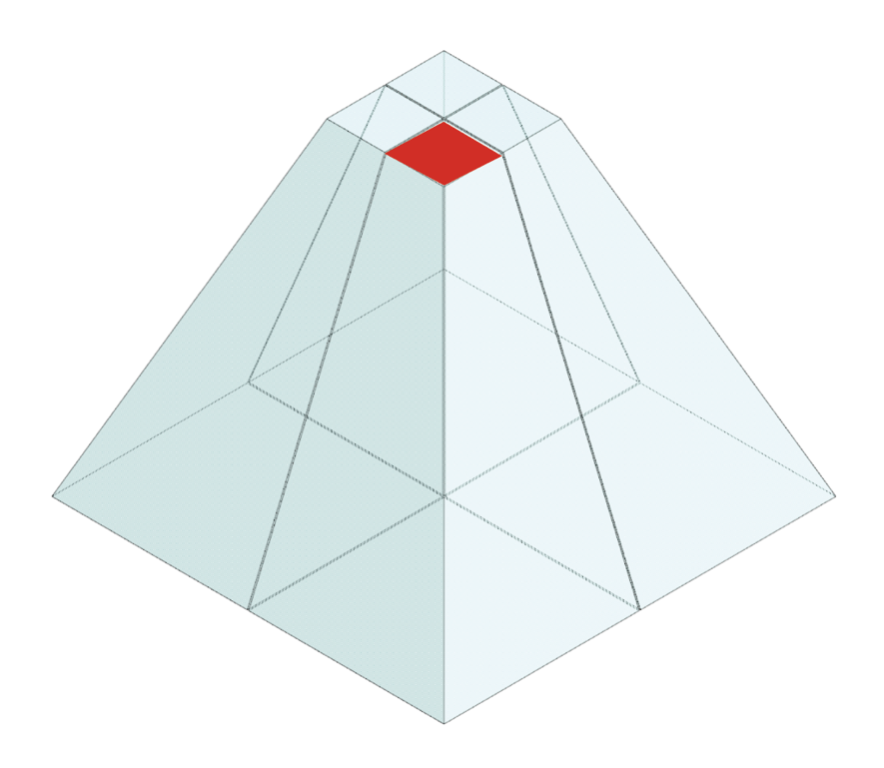}
\qquad
\includegraphics[width=.58\textwidth]{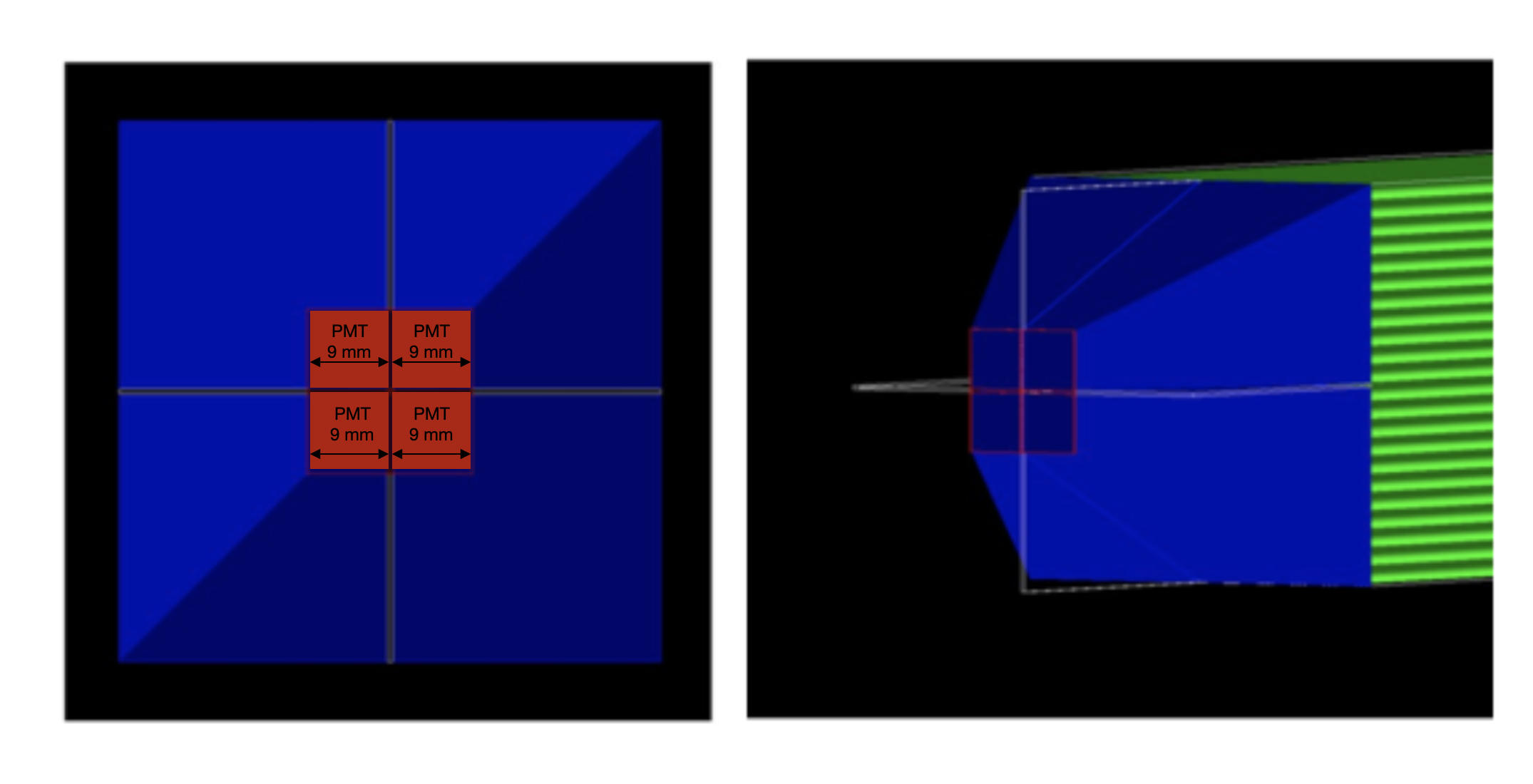}
\caption{Geometry 4: light guide geometry for a 4-cell SpaCal configuration assuming usage of a multi-anode photomultiplier (PMT).\label{fig:4LGGeo}}

\end{figure}

Several geometrical light guide configurations were studied, including symmetrical and asymmetrical trapezoidal guides and parabolic guides, commonly referred to as Winston-cones, with square and round bases, as illustrated in Figure~\ref{fig:LGGeo}. The light guide is defined by two fixed parameters: the cell size, which is fixed to $30.25\times 30.25$~mm$^2$, and the sensitive area of the photomultiplier: $18\times18$~mm$^2$ (e.g. R7600), $9\times9$~mm$^2$
(multi-anode version, e.g. R7600-M4, Figure~\ref{fig:4LGGeo}), or round photocathode $\oslash$8~mm (e.g. R9880) and one variable parameter - length which is considered in the range from 20~mm to 50~mm with 10~mm step.

A reflective mirror was installed at the back of each SpaCal cell to create a single readout scenario and account for all scintillation photons.

\section{Gaussian approximation for the light collection efficiency}
\label{sec:GaussianApprox}
\begin{figure}[htbp]
\centering
\includegraphics[width=.46\textwidth]{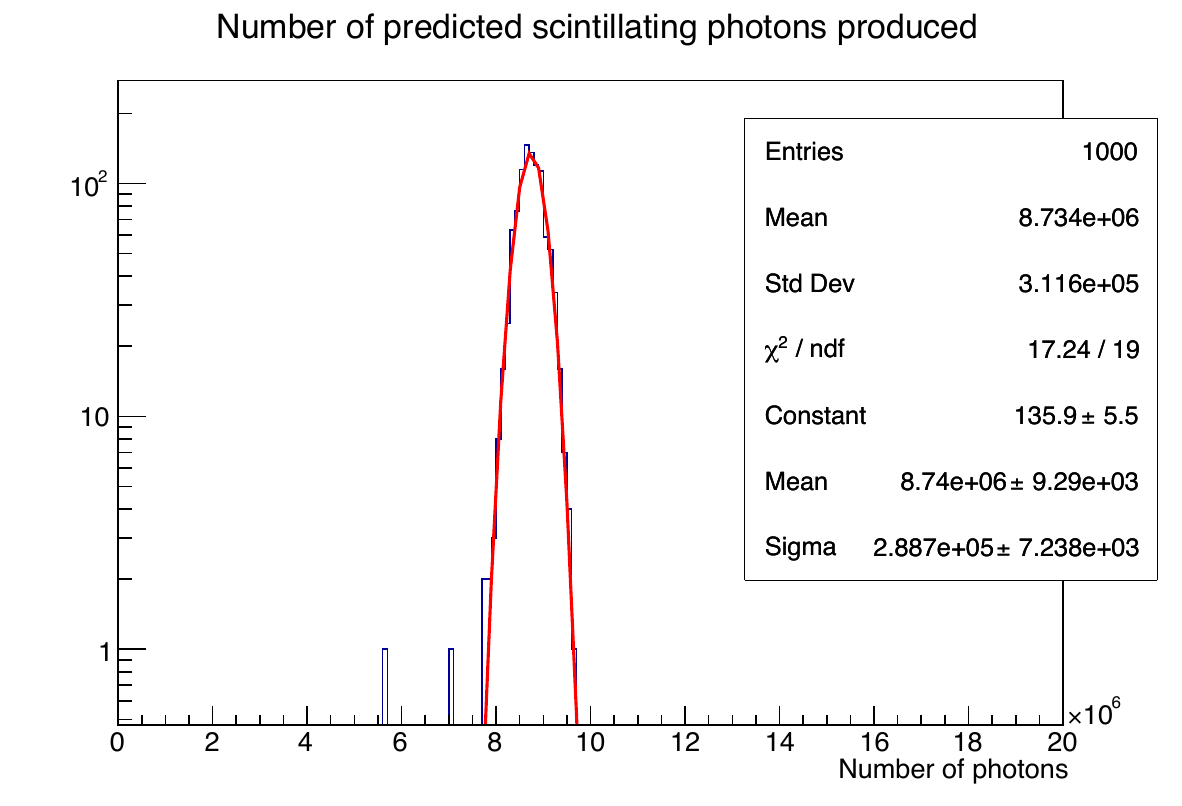}
\qquad
\includegraphics[width=.46\textwidth]{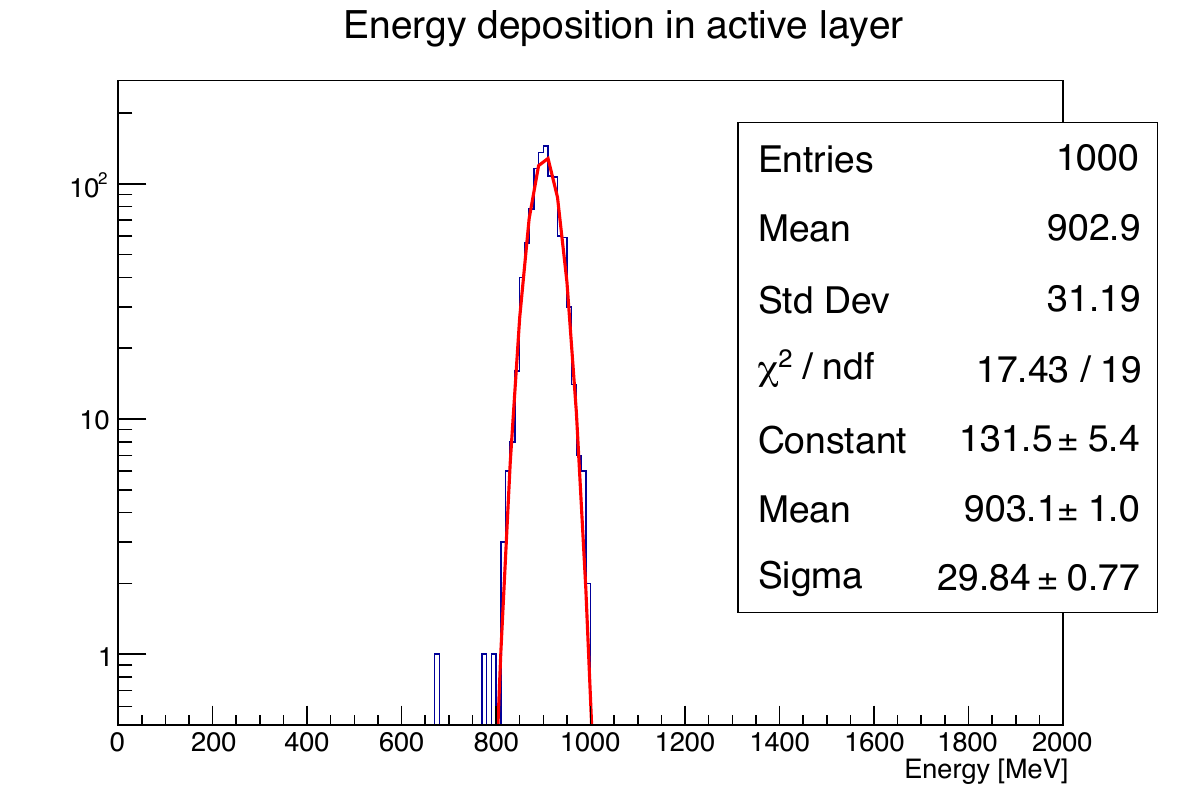}
\caption{Predicted number of scintillating photons produced~(left) and energy deposit in the active material~(right) by the electrons with 10~GeV energy at a \(3^\circ + 3^\circ\) incidence angle.}
\label{fig:edepphotons}
\end{figure}

To primarily test the impact of light collection non-uniformity on energy resolution, electrons with energies in the range from 1~GeV to 100~GeV impinging at the center of the SpaCal module at the angle of $3^\circ$ in both the horizontal and vertical planes (denoted as $3^\circ+3^\circ$) are simulated. The light collection configuration is described in Section~\ref{sec:Calo}. The amount of energy deposited in each scintillating fibre is used to estimate the number of scintillation photons produced, assuming a light yield of 10,000 photons / MeV and taking into account Birks' Law.  In our simulations, the Birks’ constant $k_B$ is set to \(\text{0.126\, mm / MeV}\). Figure~\ref{fig:edepphotons} (left) shows the distribution of the predicted number of scintillation photons produced by 10~GeV electrons. Figure~\ref{fig:edepphotons} (right) shows the distribution of the energy deposited in the scintillating fibres for the same configuration. The reference energy resolution value is obtained as a simple ratio of the standard deviation and mean values of the distribution of the number of photons. 

\begin{figure}[htbp]
\begin{minipage}{0.49\linewidth}
\includegraphics[width=\linewidth]{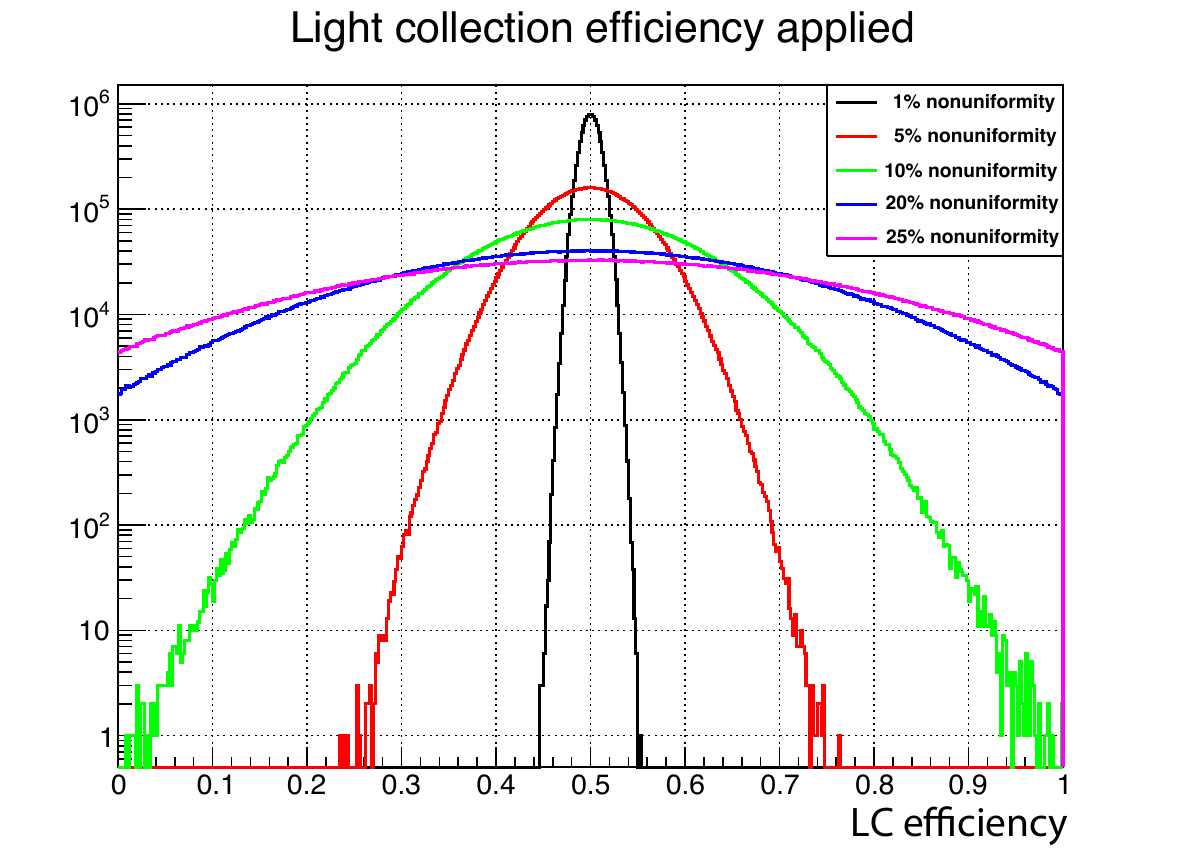}
\end{minipage}
\begin{minipage}{0.49\linewidth}
\includegraphics[width=\linewidth]{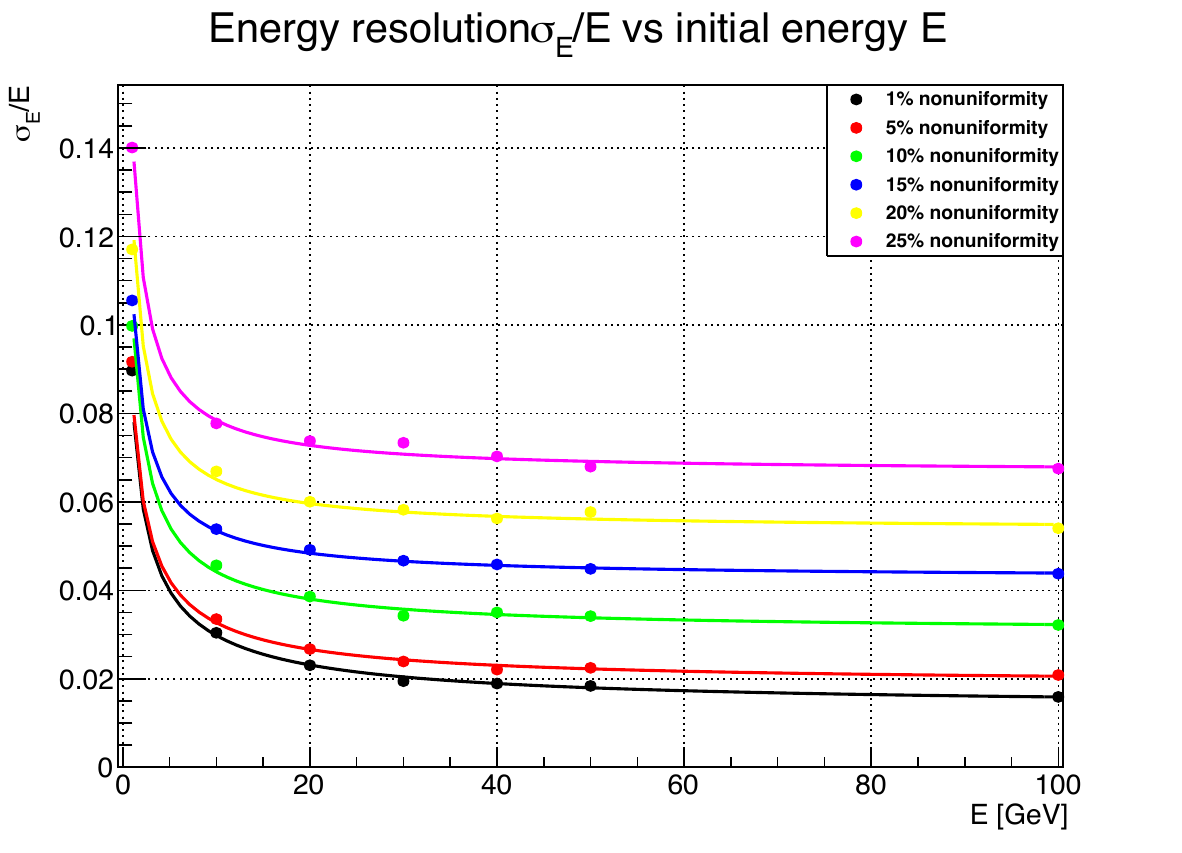}
\end{minipage}
\caption{Left: Distributions of light collection efficiencies obtained with a Gaussian distribution (mean = 0.5) and \(\sigma\) values ranging from 0.01 to 0.25, truncated to the \([0,1]\) interval. Right: Dependence of the energy resolution ($\sigma_E/E$) as a function of the incident electron energy for different levels of light collection non-uniformity.}
\label{fig:appliedEfficiency}
\end{figure}

For each event, the energy deposit in each scintillating fibre is weighted with an additional coefficient randomly obtained from the Gaussian distribution as shown in Figure~\ref{fig:appliedEfficiency} (left). The dependence of the obtained energy resolution ($\sigma_E/E$) as a function of the incident electron energy is shown in Figure~\ref{fig:appliedEfficiency} (right). Different colours are used to present different $\sigma$ parameter values for non-uniformity coefficients.

\begin{figure}[htbp]
\centering
\includegraphics[width=.75\linewidth]{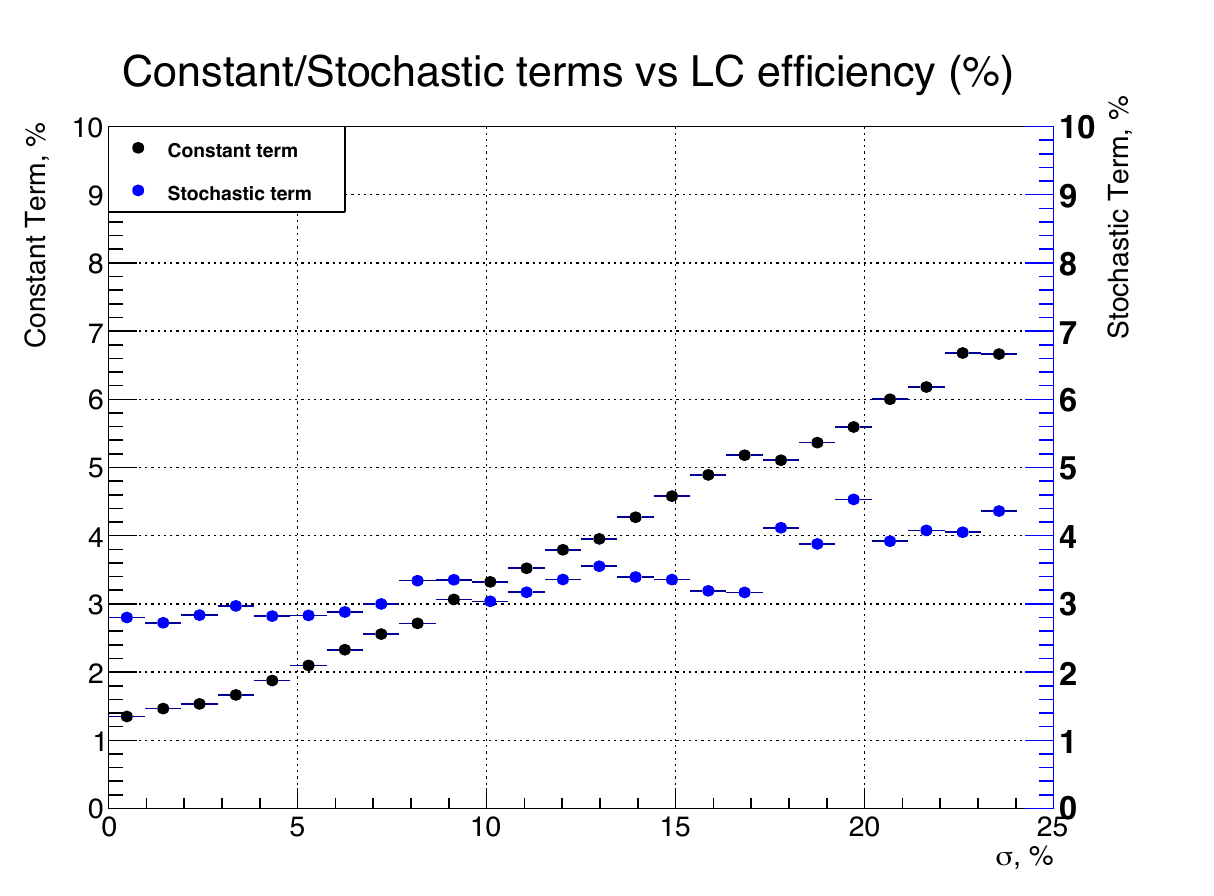}
\caption{Constant (black) and stochastic (blue) terms of the energy resolution for electrons as a function of the $\sigma$ value of the Gaussian distribution used to estimate the light collection efficiency.}
\label{fig:GausER}
\end{figure}

The two terms~(constant and stochastic) of the energy resolution as a function of the incoming particle energy are obtained and presented in Figure~\ref{fig:GausER} as a function of the $\sigma$ of a sigma value used to generate the light collection coefficient distribution. As expected, the increase in the $\sigma$ parameter leads to a clear increase in the constant terms.


\section{Optical simulation to estimate the efficiency of the light guide}
\label{sec:EfficiencyLG}

To evaluate the light collection efficiency of the light guide, muons with a momentum of 1~GeV/$c$ were simulated. Particles entered the test module perpendicularly to the fibre direction (i.e., across the fibre axis). Muons were selected for this study because they deposit energy uniformly along their trajectory without generating electromagnetic showers, enabling an assessment of light collection efficiency without interference from shower development. In this evaluation, efficiency denotes the relative light collection efficiency of each fiber, defined as the fraction of scintillation photons that enter the light guide and successfully reach the photodetector (PMT). For each fiber, scintillation photons were generated within the fiber and tracked throughout their propagation. We recorded the total number of photons produced and the subset that reached the photodetector. All key optical properties (e.g., refractive indices, absorption lengths, and surface finishes) were included.




The different light guide geometries under study included those described in Section~\ref{sec:Calo}, with light guide lengths ranging from 20~mm to 50~mm in increments of 10~mm. Air was used as the LG material in these simulations to focus on the effects of LG geometry, minimizing material-dependent effects.

\begin{figure}[htbp]
\centering
\begin{subfigure}{0.62\textwidth}
  \includegraphics[width=\linewidth]{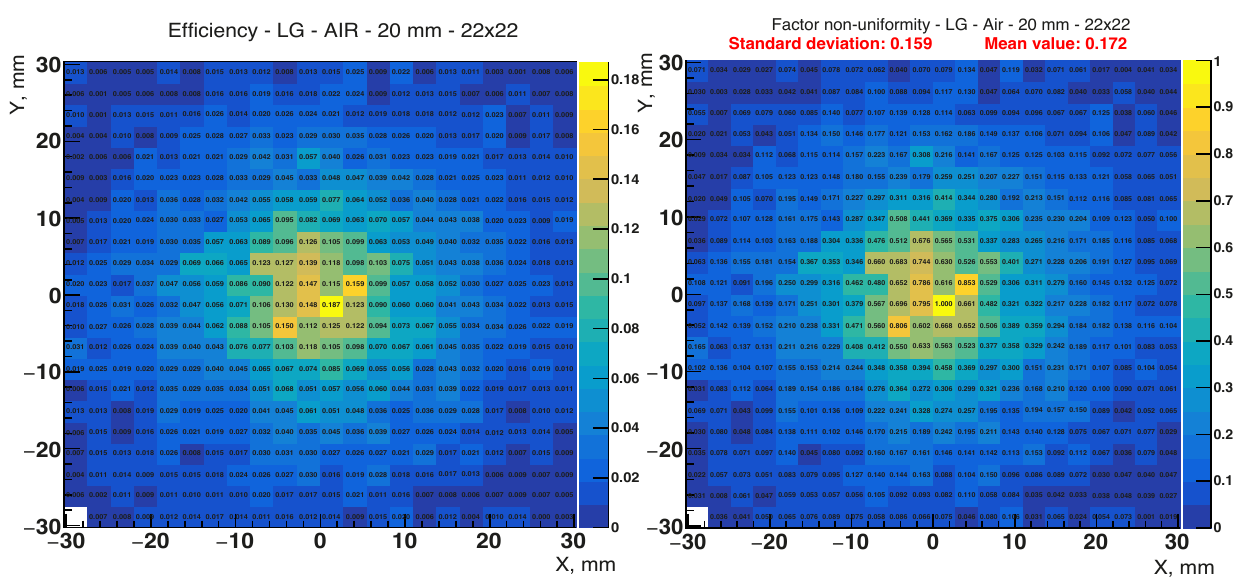}
  \caption{20 mm LG length}
\end{subfigure}
\begin{subfigure}{0.62\textwidth}
  \includegraphics[width=\linewidth]{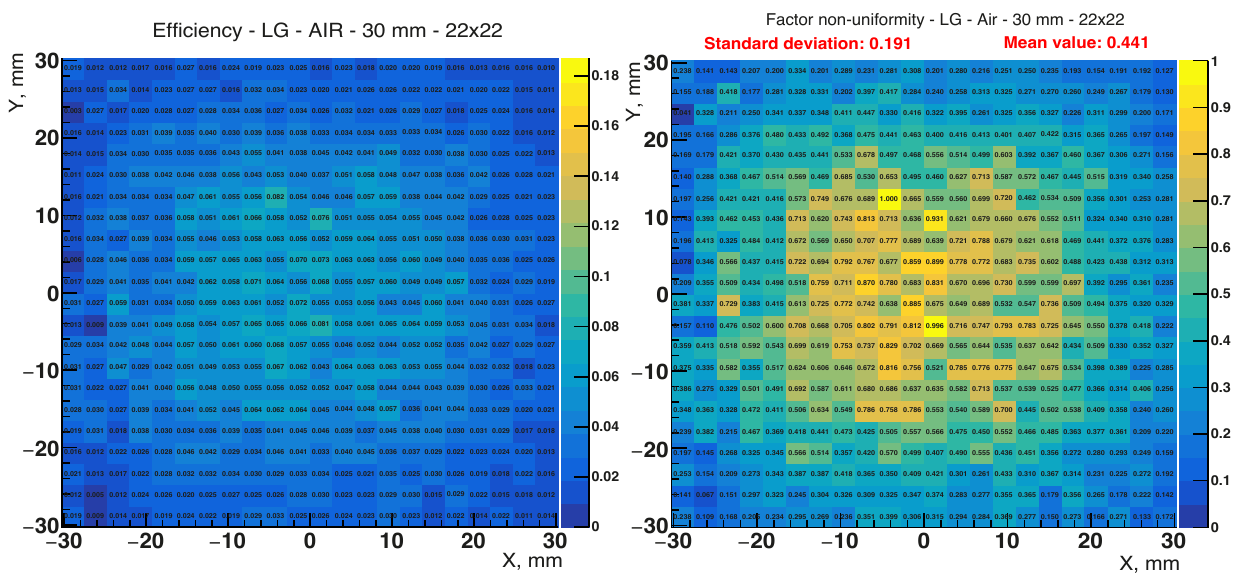}
  \caption{30 mm LG length}
\end{subfigure}\\
\begin{subfigure}{0.62\textwidth}
  \includegraphics[width=1.\linewidth]{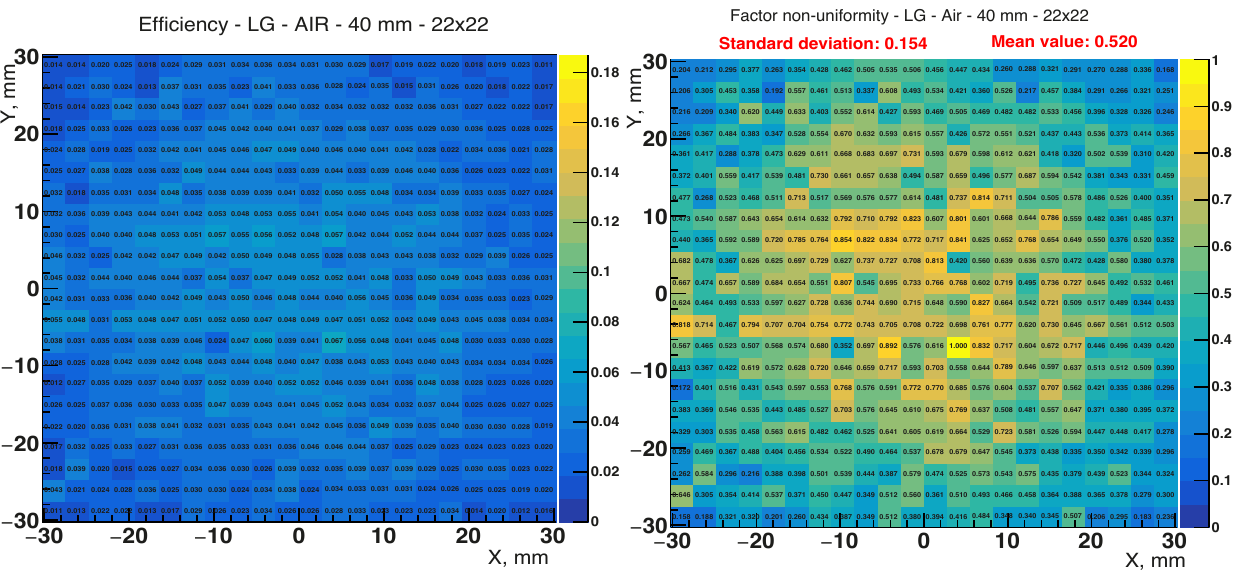}  
  \caption{40 mm LG length}
\end{subfigure}
\centering
\begin{subfigure}{0.62\textwidth}
  \includegraphics[width=1.\linewidth]{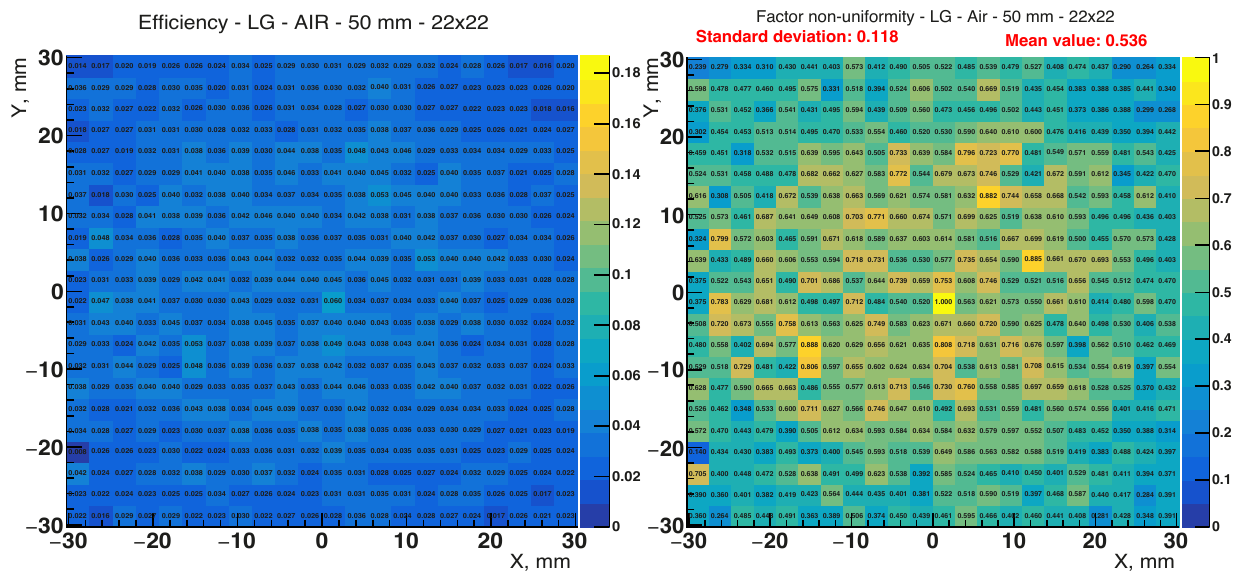}  
  \caption{50 mm LG length}
\end{subfigure}
\caption{Efficiency maps for Geometry 4 with varying light guide lengths: a) 20 mm, b) 30 mm, c) 40 mm,  d) 50 mm. Each panel consists of two subplots. The left subplot shows the individual fibre light collection efficiency in the \(xy\)-plane for a 22x22 grid of 4 SpaCal cells. The right subplot presents the distribution of the non-uniformity factor in the same \(xy\)-plane. The red text in the title shows the non-uniformity mean and standard deviation values.}
\label{fig:LG4_eff_example}
\end{figure}

Efficiency maps were generated for each light guide geometry and each length of the light guide to visualize the spatial distribution of light collection efficiency across the cell. Here, efficiency is defined as the ratio of the number of photons reaching the PMT to the number entering the light guide. These maps were obtained by scanning 1\,GeV/$c$ muons across SpaCal cells and recording the efficiency at each $(x,y)$ position. Figure~\ref{fig:LG4_eff_example} (left) shows examples for Geometry~4 with light guide lengths of 20--50\,mm. These maps represent a $60.5 \times 60.5$~mm$^2$ area, corresponding to a $2 \times 2$ arrangement of SpaCal cells. In each case,  the efficiency is normalized to the maximum fibre response value  (Figure~\ref{fig:LG4_eff_example} - right), and the factor non-uniformity refers to this ratio of local-to-maximum efficiency. The mean and standard deviation shown on each plot quantify how uniformly light is collected across the scanned area: the smaller the standard deviation, the more uniform the response. 


\begin{figure}[htbp]
\centering
\includegraphics[width=.6\textwidth]{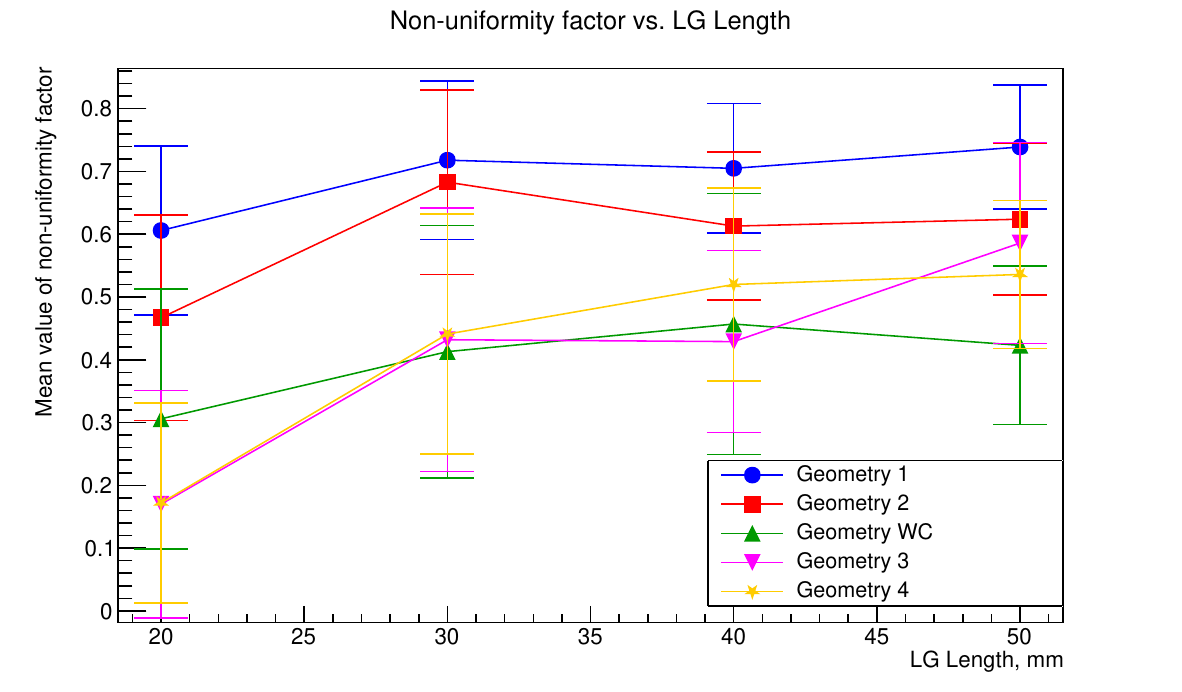}
\caption{The plot illustrates the relationship between the mean non-uniformity factor for a single SpaCal cell and the length of the LG for different geometry. Geometry 1 is shown in blue, Geometry 2 in red, Geometry WC in green, Geometry 3 in magenta, and Geometry 4 in yellow. }
\label{fig:distUnirom}
\end{figure}

As observed in Figure~\ref{fig:distUnirom}, shorter light guides exhibit a higher sigma value in the non-uniformity, indicating a larger relative variation in the light collection efficiency across the cell. Since our primary focus is on achieving a uniform light collection response rather than maximizing the total number of collected photons, designs featuring longer light guides are favored.

\section{Energy resolution for SpaCal module}
\label{sec:EnergyResolution}

\begin{figure}[htbp]
\begin{subfigure}{.49\linewidth}
  \centering
  \subfloat[\label{fig:geo1_2cm}\centering]{
  \includegraphics[clip, trim=0cm 0cm 0cm 1cm, width=1.\linewidth]{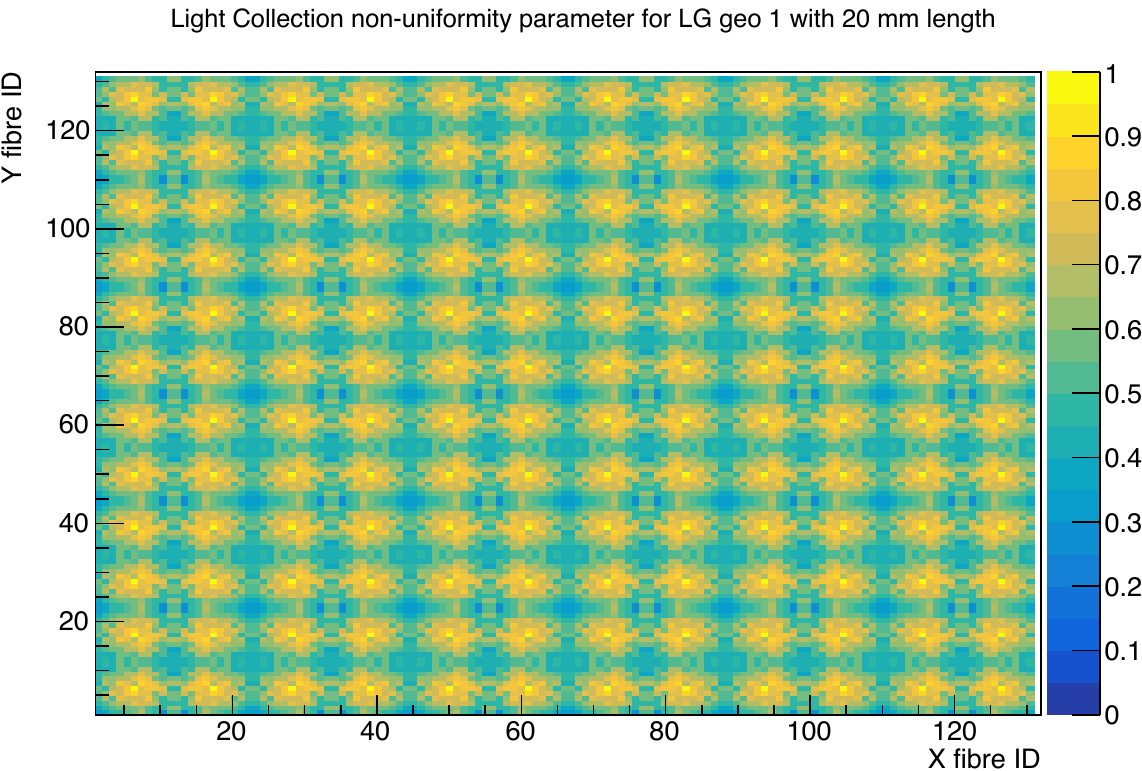}}
\end{subfigure}
\hfill
\begin{subfigure}{.49\linewidth}
  \centering
  \subfloat[\label{fig:geo1_5cm}\centering]{
  \includegraphics[clip, trim=0cm 0cm 0cm 1cm, width=1.\linewidth]{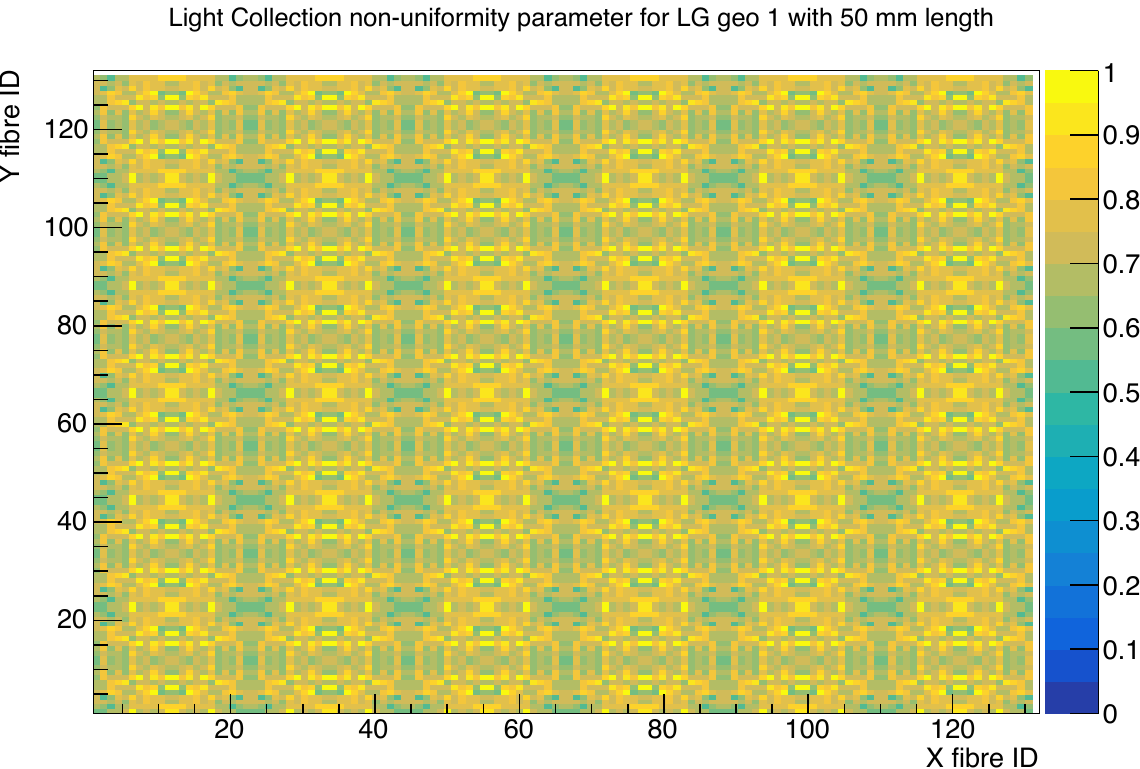}}
\end{subfigure}
\newline
\begin{subfigure}{.49\linewidth}
  \centering
  \subfloat[\label{fig:geo4_2cm}\centering]{
  \includegraphics[clip, trim=0cm 0cm 0cm 1cm, width=1.\linewidth]{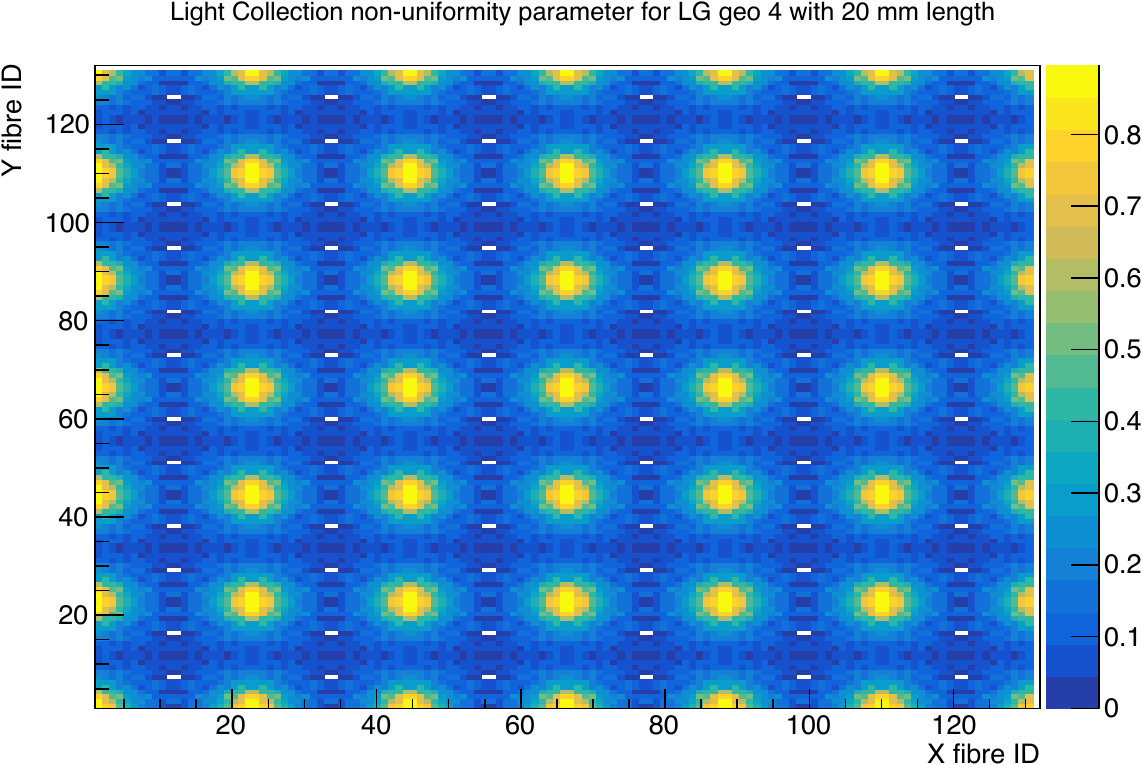}}
\end{subfigure}
\begin{subfigure}{.49\linewidth}
  \centering
  \subfloat[\label{fig:geo4_5cm}\centering]{
  \includegraphics[clip, trim=0cm 0cm 0cm 1cm, width=1.\linewidth]{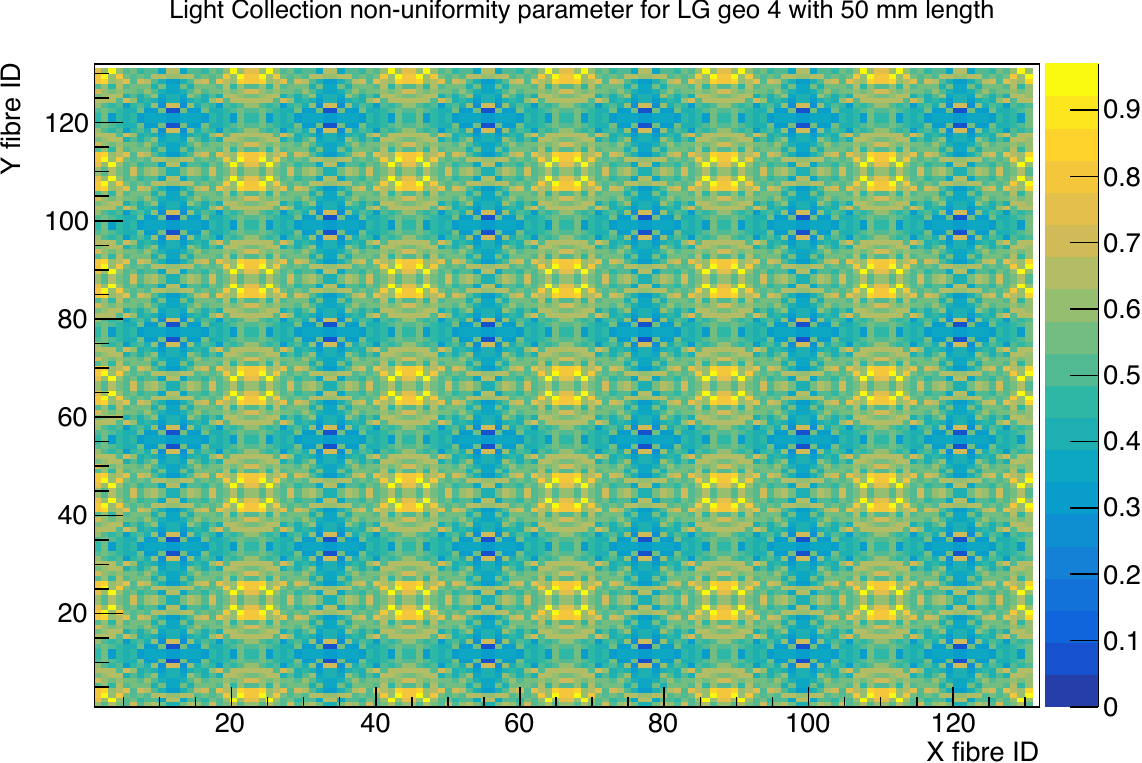}}
\end{subfigure}
\caption{
Non-uniformity factor for the SpaCal module as a function of fiber position (X fiber ID and Y fiber ID) for different LG configurations. Each subplot represents the light collection non-uniformity for a specific geometry and LG length: 
Figure~\ref{fig:geo1_2cm}: Geometry 1 with 20 mm long light guide; Figure~\ref{fig:geo1_5cm}: Geometry 1 with 50 mm long light guide; Figure~\ref{fig:geo4_2cm}: Geometry 4 with 20 mm long light guide; Figure~\ref{fig:geo4_5cm}: Geometry 4 with 50 mm long light guide.}
\label{fig:non-uniformfactor}
\end{figure}

To assess the impact of light collection non-uniformity on the overall performance of the SpaCal module, we extended our simulation from a single cell to the entire calorimeter module. Each cell contains scintillating fibres and a light guide, configured according to the geometries described in Section~\ref{sec:Calo}. Non-uniformity factors observed in the single-cell simulations were applied to each fiber in the module to create the calorimeter's performance (see Figure~\ref{fig:non-uniformfactor}).

\begin{figure}[htbp]
\begin{subfigure}{.49\linewidth}
  \centering
  \subfloat[\label{fig:geometryPerfect}\centering]{
  \includegraphics[width=\linewidth]{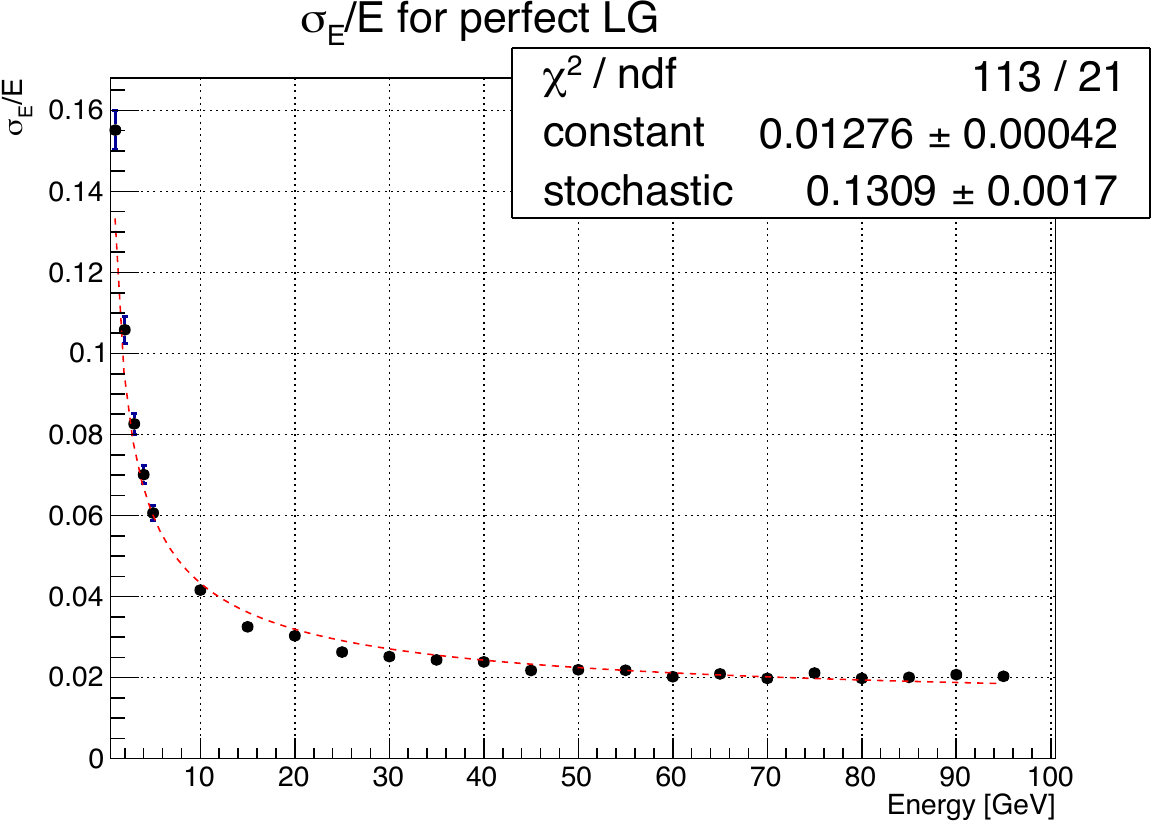}}
\end{subfigure}
\hfill
  \begin{subfigure}{.49\linewidth}
\centering
\subfloat[\label{fig:geometry1}\centering]{
  \includegraphics[width=\linewidth]{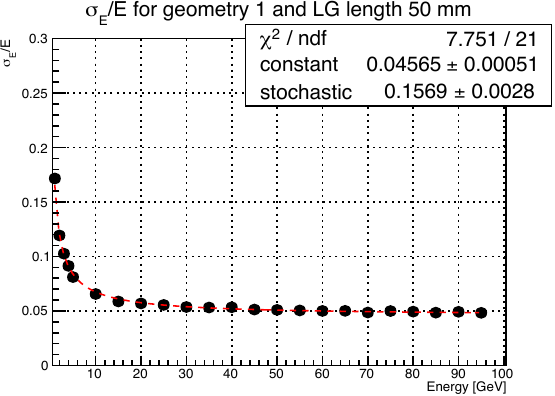}}
\end{subfigure}
\vspace{5mm}
\newline
\begin{subfigure}{.49\linewidth}
  \centering
\subfloat[\label{fig:geometry2}\centering]{
  \includegraphics[width=\linewidth]{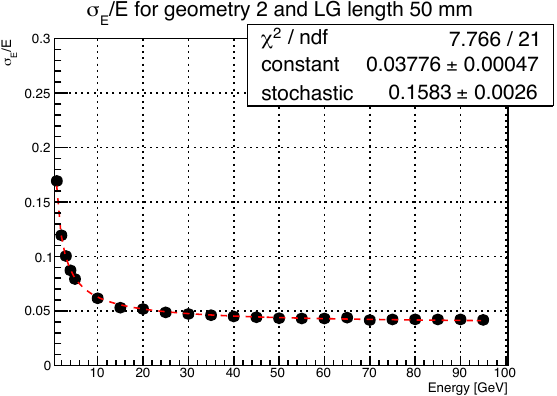}} 
\end{subfigure}
\hfill
\begin{subfigure}{.49\linewidth}
  \centering
\subfloat[\label{fig:geometryWC}\centering]{
  \includegraphics[width=\linewidth]{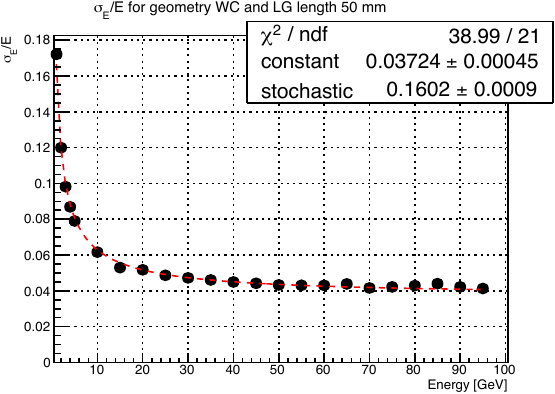}}
\end{subfigure}
\vspace{5mm}
\newline
\begin{subfigure}{.49\linewidth}
  \centering
  \subfloat[\label{fig:geometry3}\centering]{
  \includegraphics[width=\linewidth]{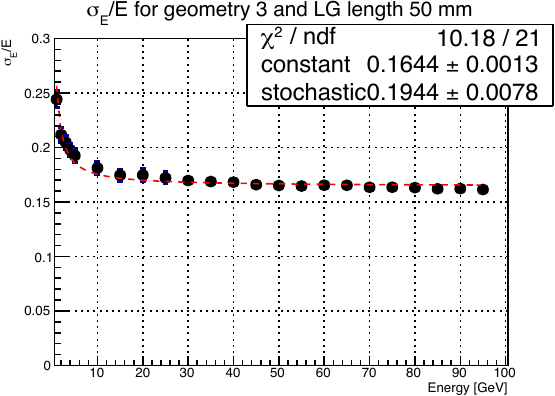}}  
\end{subfigure}
\hfill
\begin{subfigure}{.49\linewidth}
  \centering
  \subfloat[\label{fig:geometry4}\centering]{
  \includegraphics[width=1.\linewidth]{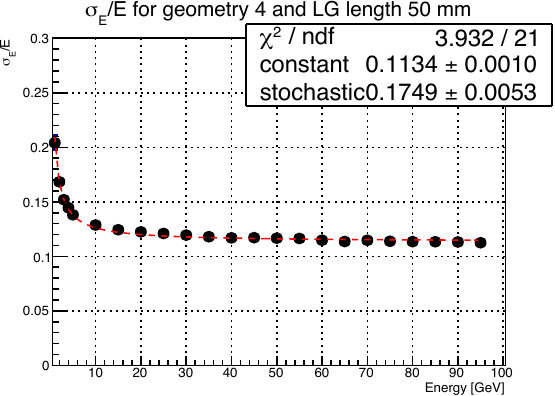}}
\end{subfigure}
\caption{
Energy resolution (\(\sigma_E / E\)) for electrons at a \(3^\circ + 3^\circ\) incidence angle across different LG configurations. Each plot presents effects coming from a specific geometry: \ref{fig:geometryPerfect} - Perfect light guide (energy deposit only); \ref{fig:geometry1} Geometry 1 with 50 mm LG; \ref{fig:geometry2} Geometry 2 with 50 mm LG; \ref{fig:geometryWC} Geometry WC with 50 mm LG; \ref{fig:geometry3} Geometry 3 with 50 mm LG; and \ref{fig:geometry4} Geometry 4 with 50 mm LG.}
\label{fig:energyresol_graph}
\end{figure}

The energy resolution was calculated for electrons of 1 to 100~GeV at 3$^{\circ}$+3$^{\circ}$ incidence angle. The energy resolution as a function of the incoming particle energy is described by a stochastic $s$ and a constant $c$ terms:
\begin{equation}
\label{eq:energyRes}
    \frac{\sigma}{E} = \frac{s}{\sqrt{E}} \oplus c,
\end{equation}
where $\oplus$ is a quadratic sum, and $E$ is the energy of the incident electron in GeV. Figure~\ref{fig:energyresol_graph} shows the energy resolution across various light guide geometries and lengths ranging from 20~mm to 50~mm. Asymmetrical geometries exhibited worse energy resolution due to increased non-uniformity in the light collection. Symmetric and parabolic geometries, such as the Winston-cone, performed better by providing more uniform light distribution across the module, thus improving energy resolution.

\begin{figure}[htbp]
\centering
\begin{subfigure}{.5\textwidth}
  \centering
  \includegraphics[width=1.\linewidth]{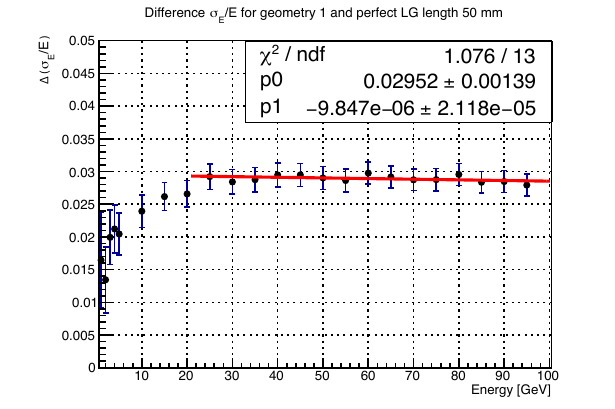}  
  \caption{Geometry 1 with 50 mm LG}
  \label{fig:sub-first}
\end{subfigure}%
\begin{subfigure}{.5\textwidth}
  \centering
  \includegraphics[width=1.\linewidth]{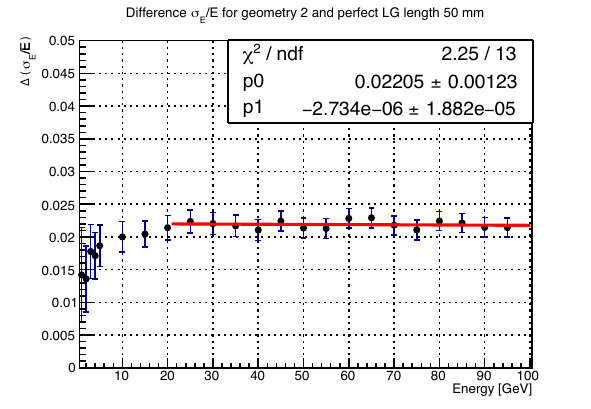}  
  \caption{Geometry 2 with 50 mm LG}
  \label{fig:sub-second}
\end{subfigure}
\newline
\begin{subfigure}{.5\textwidth}
  \centering
  \includegraphics[width=1.\linewidth]{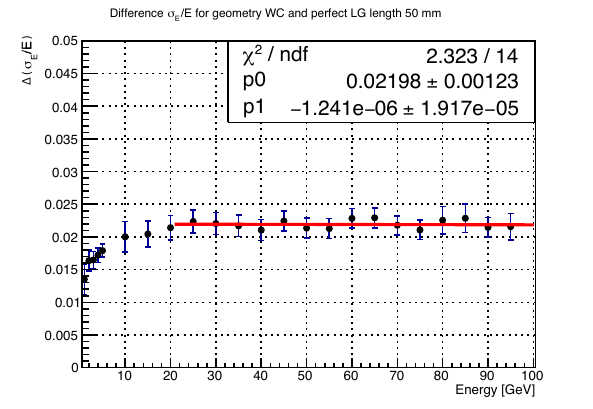}  
  \caption{Geometry WC with 50 mm LG}
  \label{fig:sub-third}
\end{subfigure}%
\begin{subfigure}{.5\textwidth}
  \centering
  \includegraphics[width=1.\linewidth]{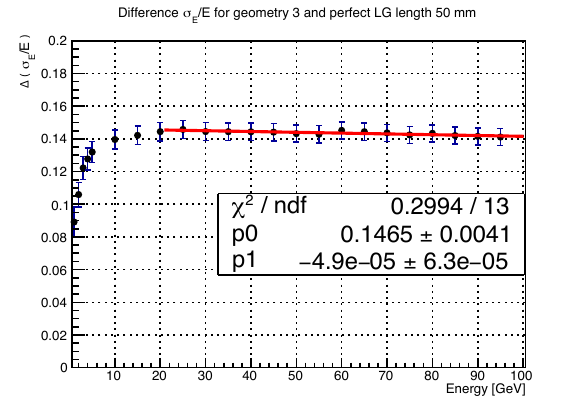}  
  \caption{Geometry 3 with 50 mm LG}
  \label{fig:sub-fourth}
\end{subfigure}
\newline
\begin{subfigure}{.5\textwidth}
  \centering
  \includegraphics[width=1.\linewidth]{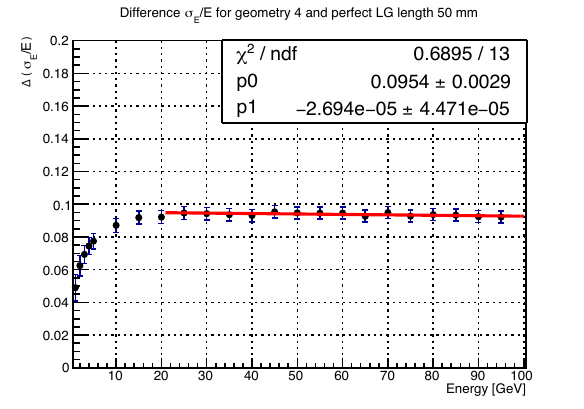}  
  \caption{Geometry 4 with 50 mm LG}
  \label{fig:sub-fifth}
\end{subfigure}
\caption{
Difference in energy resolution \(\Delta\)(\( \sigma_E / E\)) between different LG geometries and the perfect LG configuration with a uniform light collection. Each subfigure shows the absolute difference in energy resolution as a function of energy for a specific geometry (Geometry 1, Geometry 2, Geometry WC, Geometry 3 and Geometry 4), compared to the perfect configuration. A linear fit is applied to quantify the trend of energy resolution deviation across the energy range (from 20 GeV). Lower values of \(\Delta\)(\( \sigma_E / E\)) indicate closer alignment with the ideal configuration. 
}
\label{fig:Perfect_delta}
\end{figure}

The best result was obtained for an ideal light guide, with uniform light collection from each fibre (Figure~\ref{fig:geometryPerfect}). 
The contribution from the light guide should appear as an extra term in the Eq.~\ref{eq:energyRes}. Thus could be obtained after substructing quadratically from the energy resolution with considered light guide geometry the corresponding value obtained with energy deposit only (reference value). Figure~\ref{fig:Perfect_delta} shows the obtained difference \(\Delta\)(\(\sigma_E / E\)) as a function of incoming particle energy. 
The numerical values of the results are summarized in Table~\ref{tab:i}.

\begin{table}[htbp]
\centering
\smallskip
\begin{tabular}{c|c|c|c|c}
 & \multicolumn{4}{c}{LightGuide length} \\

 & 20 mm & 30 mm & 40 mm & 50 mm \\
\hline
\hline
Geometry 1 & \multicolumn{4}{c}{} \\
\hline
Const term & 0.106 $\pm$ 0.001  & 0.073 $\pm$ 0.001 & 0.050 $\pm$ 0.001 & 0.046 $\pm$ 0.001  \\
Stoch term & 0.202 $\pm$ 0.005 & 0.175 $\pm$ 0.004 & 0.162  $\pm$ 0.003 & 0.157 $\pm$ 0.003\\ 
\hline
Geometry 2 & \multicolumn{4}{c}{} \\
\hline
Const term  & 0.151 $\pm$ 0.001 & 0.064 $\pm$ 0.001 & 0.039 $\pm$ 0.001 & 0.038 $\pm$ 0.001 \\
Stoch term & 0.248 $\pm$ 0.007 & 0.173 $\pm$ 0.003 & 0.160 $\pm$ 0.003 & 0.158 $\pm$ 0.003\\
\hline
Geometry WC & \multicolumn{4}{c}{} \\
\hline
Const term  & 0.157 $\pm$ 0.001 & 0.064 $\pm$ 0.001 & 0.039 $\pm$ 0.001 & 0.037 $\pm$ 0.001 \\
Stoch term & 0.231 $\pm$ 0.003 & 0.172 $\pm$ 0.001 & 0.162 $\pm$ 0.001 & 0.160 $\pm$ 0.001\\
\hline
Geometry 3 & \multicolumn{4}{c}{} \\

\hline
Const term & 0.768 $\pm$  0.006 & 0.334 $\pm$ 0.003 & 0.216  $\pm$ 0.002 & 0.164 $\pm$ 0.001\\
Stoch term & 0.495 $\pm$ 0.045 & 0.279 $\pm$ 0.017 & 0.220 $\pm$ 0.010 & 0.194 $\pm$ 0.008\\
\hline
Geometry 4 & \multicolumn{4}{c}{} \\
\hline
Const term  & 0.674 $\pm$ 0.005 & 0.307 $\pm$ 0.002 & 0.180 $\pm$ 0.001 & 0.113 $\pm$ 0.001\\
Stoch term & 0.511 $\pm$ 0.036 & 0.260 $\pm$ 0.016 & 0.203 $\pm$ 0.009 & 0.175 $\pm$ 0.005\\
\hline
\end{tabular}
\caption{Constant and stochastic terms of the energy resolution obtained for various geometries and lengths of the light guides.}
\label{tab:i}
\end{table}

\begin{figure}[htbp]
\centering
\includegraphics[width=1.\textwidth]{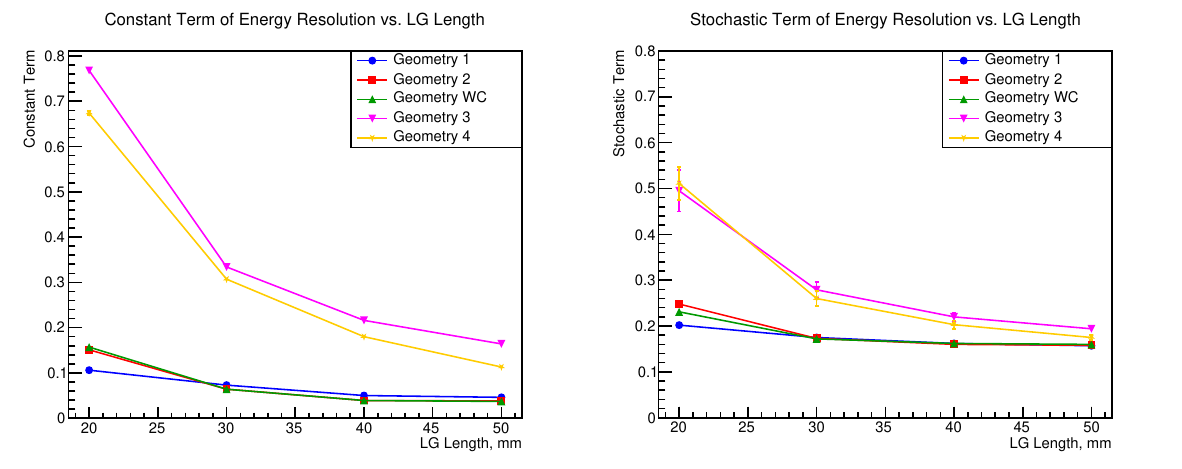}
\caption{
The plots show the dependence of the constant term (left) and the stochastic term (right) of the energy resolution on the LG length for different geometry. Geometry 1 is shown in blue, Geometry 2 in red, Geometry WC in green, Geometry 3 in magenta, and Geometry 4 in yellow. }
\label{fig:FoMPlots}
\end{figure}

By analyzing both the constant and stochastic terms as a function of LG length for different geometries (see Figure~\ref{fig:FoMPlots}), the influence of LG design choices (geometry and length) on the SpaCal module’s overall energy resolution is illustrated.

\section{Conclusion}
Non-uniformities in the light-collection lead to the additional term in the energy resolution. This addition increases with the increase of the incoming particle energy and becomes constant for energies above 20 GeV. Behaviour strongly depends on the light guide geometry, with a strong preference for the symmetric lightguide geometry and length of 40 mm and above.

This study highlights the importance of both the initial light-guide design and quality assurance measures required during the production to keep the energy resolution at the desired level.

\acknowledgments
We thank our colleagues from the LHCb ECAL Upgrade Phase 2 working group for their valuable feedback.


\begin{thebibliography}{99}

\bibitem{Scheel}
Scheel, C.V. (Dec 1994).
The spaghetti calorimeter. Research, development, application.
\url{http://inis.iaea.org/search/search.aspx?orig_q=RN:27032174}

\bibitem{Armstrong:1998qs}
T.~A.~Armstrong, K.~Barish, S.~J.~Bennett, T.~M.~Cormier, R.~Cernej, A.~Chikanian, S.~D.~Coe, R.~Davies, P.~R.~Dee and G.~E.~Diebold, \textit{et al.}
``The E864 lead-scintillating fiber hadronic calorimeter,''
Nucl. Instrum. Meth. A \textbf{406} (1998), 227-258
doi:10.1016/S0168-9002(98)91984-2


\bibitem{Kholodenko:2022anr}
S.~Kholodenko [LHCb ECAL UII working group],
``LHCb ECAL upgrade II,''
PoS \textbf{PANIC2021} (2022), 100
doi:10.22323/1.380.0100


\bibitem{Shmanin:2023oqo}
E.~Shmanin,
``Simulation Studies of the Lead-Polystyrene SPACAL Prototype for the LHCb ECAL Upgrade II,''
Phys. Atom. Nucl. \textbf{86} (2023) n o.6, 1444-1449
doi:10.1134/S1063778824010496

\bibitem{Lindner:2866493}
CERN-LHCC-2023-005, LHCB-TDR-024,
https://cds.cern.ch/record/2866493

\bibitem{GEANT4:2002zbu}
S.~Agostinelli \textit{et al.} [GEANT4],
Nucl. Instrum. Meth. A \textbf{506} (2003), 250-303
doi:10.1016/S0168-9002(03)01368-8


\bibitem{Allison:2006ve}
J.~Allison, K.~Amako, J.~Apostolakis, H.~Araujo, P.~A.~Dubois, M.~Asai, G.~Barrand, R.~Capra, S.~Chauvie and R.~Chytracek, \textit{et al.}
IEEE Trans. Nucl. Sci. \textbf{53} (2006), 270
doi:10.1109/TNS.2006.869826

\bibitem{Allison:2016lfl}
J.~Allison, J.~Apostolakis, S.~B.~Lee, K.~Amako, S.~Chauvie, A.~Mantero, J.~I.~Shin, T.~Toshito, P.~R.~Truscott and T.~Yamashita, \textit{et al.}
Nucl. Instrum. Meth. A \textbf{835} (2016), 186-225
doi:10.1016/j.nima.2016.06.125


\bibitem{kuraray}
Kuraray, Plastic scintillating fibres,
\url{https://www.kuraray.com/uploads/5a717515df6f5/PR0150_psf01.pdf}


\end{thebibliography}
\end{document}